\documentclass[twocolumn,showpacs,preprintnumbers,amsmath,amssymb]{revtex4}
\usepackage{tabularx,graphicx}

\usepackage{color}
\usepackage{hyperref}
\hypersetup{
    colorlinks=true,
    linkcolor=blue,
    filecolor=blue,      
    urlcolor=blue,
}

\begin{document}

\newcommand{\beq}{\begin{equation}}
\newcommand{\eeq}{\end{equation}}
\newcommand{\beqn}{\begin{eqnarray}}
\newcommand{\eeqn}{\end{eqnarray}}
\newcommand{\bmath}{\begin{subequations}}
\newcommand{\emath}{\end{subequations}}
\newcommand{\bra}[1]{\langle #1|}
\newcommand{\ket}[1]{|#1\rangle}

\title{Why only hole conductors can be superconductors}
\author{J. E. Hirsch }
\address{Department of Physics, University of California, San Diego,
La Jolla, CA 92093-0319}

\begin{abstract} 
The conventional theory of superconductivity says that charge carriers in a metal that
becomes superconducting can be either electrons or holes. I argue that this is incorrect. In order to satisfy conservation of 
mechanical momentum and of entropy of the universe in the superconductor to normal transition in the presence of a
magnetic field it is necessary that the normal state charge carriers are holes.
 I will
also review the empirical evidence in favor of the hypothesis that all superconductors are
hole superconductors, and discuss the implications of this for the search for higher $T_c$ 
superconductors. \end{abstract}
\maketitle
 \section{introduction}
We call the charge carriers `electrons' when the Fermi level is close to the bottom of the electronic energy band, and
`holes' when the Fermi level is close to the top of the band. Of course in metals with complicated band structures there will
be both electron carriers and hole carriers. 
 The concept of holes in   solids was introduced by  Heisenberg  \cite{heisenberg}
 and Peierls \cite{peierls}.  Upon the suggestion of Heisenberg, Peierls \cite{peierlshall} showed that holes explain the anomalous (positive) Hall coefficient of many metals. For a more recent discussion on electrons and holes, see Ashcroft and Mermin \cite{am}.
 
In the early days of superconductivity, several researchers pointed out that there appeared to be a relation between the sign
 of the Hall coefficient and superconductivity \cite{h1,h2,h3,h4,h5,h6}.
In particular that superconductivity is favored by a positive Hall coefficient, i.e. hole carriers. 
However, no explanation for this correlation was proposed.
The concept fell out of favor because it is not part of the conventional BCS-London theory of superconductivity \cite{londonbook,tinkham},
 for which electron and hole
carriers are completely equivalent.

The essential difference between electrons and holes is that electrons are deflected by a magnetic field in the direction of the
magnetic Lorentz force acting on negative carriers, and holes are deflected by a magnetic field in  the direction of  the
magnetic Lorentz force acting on positive carriers. The first situation occurs in metals with negative Hall coefficient, where we call the
carriers `electrons',  the second occurs in metals
with positive Hall coefficient, where we call the carriers `holes'. It may appear that this is not a difference but  instead reflects a fundamental
``electron-hole symmetry''. Nothing could be further from the truth.

The point is, the mobile particles in metals are always electrons, which
are negatively charged particles, whether the metal has negative or positive Hall coefficient. This has a concrete physical meaning.
It means that  when a current flows in a metal, the mechanical momentum associated with this current has opposite direction to the current itself.
Mathematically, 
\beq
\vec{\mathcal{P}}(\vec{r})=\frac{m_e}{e}\vec{J}(\vec{r})
\eeq
where $\vec{J}(\vec{r})$ is the current density at position $\vec{r}$, $\vec{\mathcal{P}}(\vec{r})$ is the mechanical momentum density 
at position $\vec{r}$, and $m_e$ and $e$ ($<0$) are the electron's bare mass and charge. Eq. (1) is valid for any band filling and band structure,
whether the Hall coefficient is negative or positive or zero. 
`Holes' are not ``real'' positive charge carriers carrying positive momentum, they are just a theoretical construct. 
This very basic electron-hole {\bf \it a}symmetry,  we argue, is the ultimate reason why only hole conductors can be superconductors.

The  reason is  simply explained in plain words, no equations are needed. 
If a moving electron is deflected in direction opposite to what is dictated by the magnetic Lorentz force acting on it, as happens in
metals with positive Hall coefficient, it is not because the sign of the electron charge has magically changed. It is because another force is acting on the electron and is dominant. This other force is a force
exerted by the ions on the electron. In metals with hole carriers it plays a dominant role, in metals with electron carriers it doesn't, since
the electron is deflected by the magnetic Lorentz force in the direction dictated by that force.

If an electron-ion force is acting, momentum is transferred from the ions to the electrons, and by Newton's third law, {\it momentum is
transferred from the electrons to the ions}. This momentum transfer occurs in a reversible fashion, without irreversible scattering processes.
It occurs because the electron-ion interaction becomes dominant for carriers near the top of bands, where the de Broglie wavelength
of the carrier becomes comparable to the interatomic spacing. Instead, carriers near the bottom of bands have de Broglie wavelengts
much larger than the interatomic spacing and don't `see' the discrete nature of the electron-ion potential.

When a superconductor carrying a supercurrent goes normal, the supercurrent stops. The kinetic energy of the supercurrent is stored in
the electronic degrees of freedom, available to be used in the reverse transformation from normal to superconductor. 
The mechanical momentum of the supercurrent is transferred to the body as a whole in a reversible
fashion,  without  irreversible collisions. We argue that only hole carriers can do this for the reason given in the preceding paragraph. 

The concept of holes has played a prominent role  in semiconductor physics for a long time, as exemplified by the title of Shockley's 1950 book
``Electrons and holes in semiconductors''.  But it had played 
essentially no 
role in superconductivity until the discovery of high $T_c$ cuprates in 1986. 
Since the discovery of high $T_c$ cuprates, superconductivity researchers have paid attention to the question of whether charge carriers in the
normal state are electrons or holes \cite{muller}, or both. In many so-called `unconventional' superconductors the charge carriers are clearly holes,
in many others the situation is not clear, and there is no general agreement that this is an important question. Instead, since 1989  \cite{holefirst} we have
proposed that superconductivity can only occur in materials where hole carriers exist, and is driven by pairing of hole carriers \cite{holesc}.

Our initial motivation for this proposal was that  a hole causes a large disruption in its environment when it propagates, while an electron causes little disruption. This 
makes it favorable for holes to pair. Related to this, there is an off-diagonal matrix element of  the Coulomb interaction in the presence of the periodic 
electron-ion potential that is repulsive for electrons and attractive for holes. As a consequence, holes lower their effective mass when they pair, and as a consequence lower their kinetic
energy.  When a band is close to full and hole carriers exist, carriers near the Fermi energy have high kinetic
energy, so it is natural that pairing driven by lowering of kinetic energy would occur in that regime.

In a paper from 2005 we listed many reasons why, contrary to what is 
 generally assumed,  holes are $not$ like electrons, that are relevant to
  superconductivity, as shown in Figure 1. 
  The  one discussed in this paper is highlighted.
  
      \begin{figure} [htb]
 \resizebox{8.5cm}{!}{\includegraphics[width=6cm]{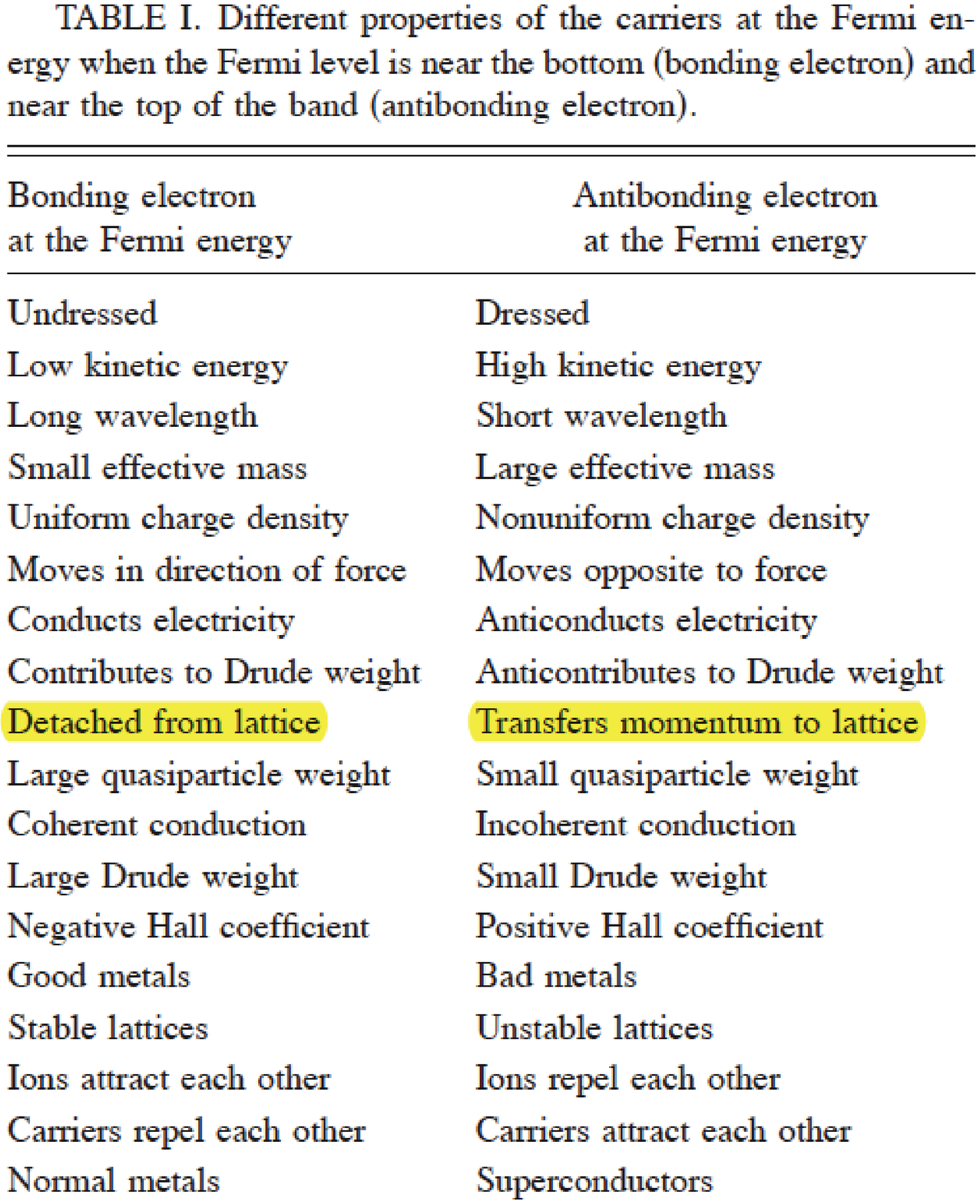}}
 \caption { From a paper the author wrote in 2005 \cite{holeelec2}. The most important reason for why
 holes are necessary for superconductivity is highlighted. }
 \label{figure1}
 \end{figure} 

 \section{Thermodynamics of the superconductor-normal transition in a magnetic field}
 The difference in free energies between normal and superconducting
  states at temperature $T<T_c$ is \cite{tinkham}
  \beq
 F_n(T)-F_s(T)=\frac{H_c(T)^2}{8\pi}
 \eeq
 where $H_c(T)$ is the thermodynamic critical field  and the free energies are given
 per unit volume. From Eq. (2) it follows that the entropy difference between normal and
 superconducting phases is ($S=-\partial F/\partial T)$ is
 \beq
 S_n(T)-S_s(T)=   \frac{L(T)}{T}=-
 \frac{d}{dT}(H_c(T)^2)  .
 \eeq
and the heat capacities in the normal and superconducting states per unit volume are related by
\bmath
\beq
C_s(T)-C_n(T)=\frac{1}{4\pi}T[(\frac{\partial H_c}{\partial T})^2+H_c\frac{\partial^2H_c}{\partial T^2}]
\eeq
and in particular at the critical temperature
 \beq
C_s(T_c)-C_n(T_c)=\frac{1}{4\pi}T_c(\frac{\partial H_c}{\partial T})^2 _{T=T_c}
\eeq
\emath
which is known as the Rutgers relation, discovered even before the Meissner effect was discovered \cite{rutgers}.

The kinetic energy density of the supercurrent at the phase boundary is 
precisely given by the difference in the free energies of normal and superconducting states Eq. (2). 
The supercurrent density  is given by
\beq
\vec{J}=en_s\vec{v}_s
\eeq
with $\vec{v}_s$  the superfluid velocity. London's equation is
\beq
\vec{\nabla}\times\vec{J}=-\frac{c}{4\pi \lambda_L^2}\vec{H}
\eeq
with $\lambda_L$ the London penetration depth. In a cylindrical geometry Eq. (6) implies
\beq
J=\frac{c}{4\pi \lambda_L}H
\eeq
and using the standard equation for the London penetration depth \cite{tinkham}
\beq
\frac{1}{\lambda_L^2}=\frac{4\pi n_s e^2}{m_e c^2}
\eeq
it follows that
\beq
v_s=\frac{e\lambda_L}{m_ec}H .
\eeq
so the kinetic energy density of the supercurrent is
\beq
K=\frac{n_s}{2}m_e v_s^2=\frac{H^2}{8\pi}
\eeq
so that from Eqs. (2) and (10)
\beq
F_n(T)-F_s(T) =K(T)
\eeq
where $K(T)$ denotes the kinetic energy density of the supercurrent at the phase boundary
where the magnetic field is $H_c(T)$.
Eq.(11) guarantees that there is phase equilibrium between the two phases \cite{londonh}.
In the superconducting phase the magnetic field and the 
supercurrent decay exponentially over the London penetration length.

      \begin{figure} [htb]
 \resizebox{8.5cm}{!}{\includegraphics[width=6cm]{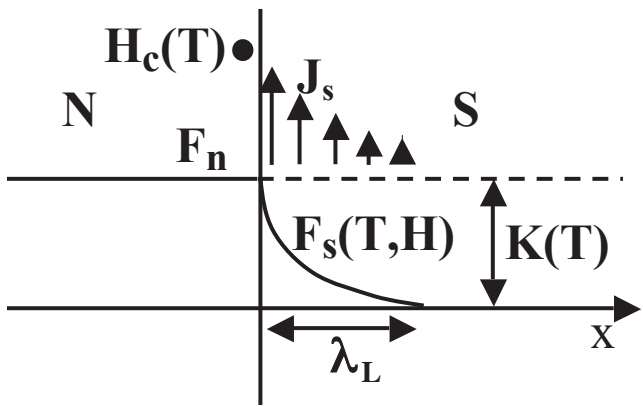}}
 \caption {Normal-superconductor (N-S) phase coexistence. 
 $F_s(T,H)$ denotes the free energy in the superconducting phase in the presence of magnetic
 field $H$, and is the sum of $F_s(T)$ and the kinetic energy of the supercurrent $K$, Eq. (10). 
 $H$ decreases from $H_c(T)$ at the phase boundary to zero 
 exponentially with characteristic length $\lambda_L$. $K(T)$ is the kinetic energy density
 of the supercurrent at the phase boundary. }
 \label{figure1}
 \end{figure}

Figure 2 shows the situation schematically. Superconducting carriers at the phase boundary have maximum
kinetic energy density  $K(T)$. The total free energy density $F_s(T,H)$  is the sum of $F_s(T)$ and the kinetic
energy density of the supercurrent for magnetic field $H$, Eq. (10).  Fig. 2 shows clearly that superconductivity is `kinetic energy driven', i.e.
associated with a lowering of electronic kinetic energy \cite{kinenergy}: as we
move into the superconducting phase, the kinetic energy of the carriers decreases and reaches its minimum deep into the 
superconducting phase.

Eq. (11)  implies that when there is a small displacement of the phase boundary whereby a region goes from S to N, or from
N to S, the resulting change in the kinetic energy of the supercurrent is exactly compensated by the difference in the
free energies of the two phases. This implies that there is zero Joule heat dissipated when
the supercurrent stops, consistent with the reversibility of the transition \cite{gorter}.

  \section{Two fluid model}
We can estimate the magnitude of various quantities involved by using the two-fluid model, which reproduces 
the properties of many superconductors   accurately \cite{shoenberg}. The critical field as function of temperature is given by
\beq
H_c(T)=H_0(1-(\frac{T}{T_c})^2)
\eeq
where $H_0$ is the thermodynamic critical field. From Eq. (12) and the thermodynamic relations given in the
previous section it follows that the entropy of the system in the superconducting and normal states is respectively
\bmath
\beq
S_s(T)=\frac{H_0^2}{2\pi T_c}(\frac{T}{T_c})^3
\eeq
\beq
S_n(T)=\frac{H_0^2}{2\pi T_c}\frac{T}{T_c}
\eeq
\emath
and as a consequence the latent heat is
\beq
L(T)=T(S_n-S_s)=\frac{H_0^2}{2\pi}(\frac{T}{T_c})^2(1-(\frac{T}{T_c})^2)
\eeq
while the kinetic energy carried by the supercurrent is
\beq
K(T)=\frac{H_c(T)^2}{8\pi}=\frac{H_0^2}{8\pi}(1-(\frac{T}{T_c})^2)^2 .
\eeq
The difference in the internal energies of the system in the normal and superconducting states is
the sum of the kinetic energy of the supercurrent and the latent heat
\beq
U_n(T)-U_s(T)=K(T)+L(T)  .
\eeq
Figure 3 shows the kinetic energy of the supercurrent, the latent heat, and their sum as function of
temperature. At low temperatures the latent heat becomes of course much smaller than the kinetic energy.

When the system goes from superconducting to normal, part of the energy required is supplied by the kinetic energy of the 
supercurrent, and the rest needs to be added as  heat in order for the temperature to remain unchanged
(otherwise the temperature would drop). The latent heat is required because
the normal system has more thermally excited states than the superconductor at given temperature. 
As the temperature approaches zero, both the entropy of the superconducting and normal states approach zero,
and the difference in internal energies is completely supplied by the kinetic energy of the supercurrent.

       \begin{figure} [htb]
 \resizebox{8.5cm}{!}{\includegraphics[width=6cm]{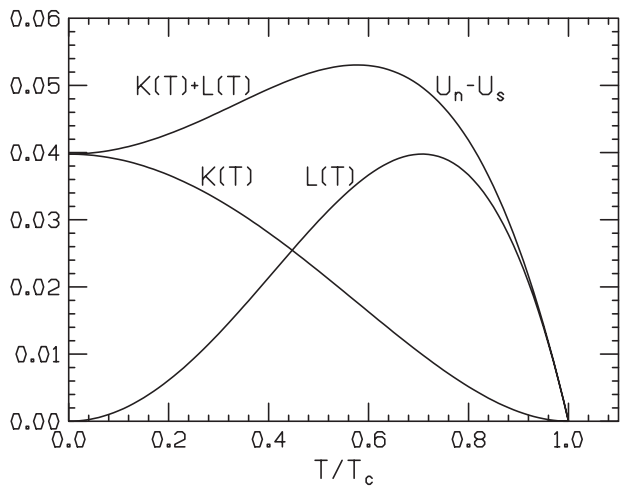}}
 \caption {Kinetic energy $K(T)$ and latent heat $L(T)$ (arbitrary units) versus temperature
 in the two-fluid model of superconductivity.
 }
 \label{figure1}
 \end{figure}

When the system goes from superconducting to normal and the supercurrent stops the kinetic energy of the
supercurrent is not dissipated as Joule heat but remains stored in the electronic normal state. This follows from
the above relations and has been verified experimentally to great accuracy \cite{keesom}. 
Particularly at low temperatures if any small fraction of the kinetic energy of the supercurrent was dissipated as
Joule heat in the transition it would be very easy to detect. Experiments  
establish that certainly not more than $1\%$ of the supercurrent kinetic energy is dissipated,
i.e. more than $99\%$ of the supercurrent is stored \cite{keesomrevers,mapother}.

 \section{can the conventional theory of superconductivity explain this?}
 
 The kinetic energy of the supercurrent is stored in the electronic degrees of freedom, but its mechanical momentum   is not: it has to be transmitted to the body as a whole to satisfy momentum conservation.
 
 Within the conventional theory of superconductivity, it is assumed that when a region goes from 
 superconducting to normal, Cooper pairs dissociate and the mechanical center of mass momentum of Cooper pairs of the supercurrent ($q$)
 is inherited by the normal electrons that are product of the dissociation ($q/2$ each) \cite{eilen}. There is no explanation in the literature of
 conventional superconductivity for how this momentum is then transmitted to the body as a whole in a reversible fashion.

 There are two questions that need to be answered within the conventional theory. First, how do the normal electrons product of dissociation of Cooper pairs
 inherit the momentum but not the kinetic energy of the Cooper pairs? Second, how do the normal electrons transfer this 
 momentum to the body as a whole without an increase in the entropy of the universe, as is required from the fact that the
 transition is reversible? Neither of these questions has been addressed in the superconductivity literature, 
 and we argue that neither of these questions has a satisfactory answer.
 
  We have argued that it is impossible to describe this physics within the conventional theory of superconductivity without violating other
 basic laws of physics, either momentum conservation, or energy conservation, or Faraday's law, or the second law of thermodynamics \cite{lorentz,lenz,missing,dyn1,dyn2,revers,disapp,momentum}.

  \section{How momentum is transferred from charge carriers to the body in a reversible fashion}
  
  The velocity of Bloch electrons is given by
\beq
\vec{v}_k=\frac{1}{\hbar}\frac{\partial \epsilon_k}{\partial \vec{k}}
\eeq
and the acceleration by
\beq
\frac{d \vec{v}_k}{dt}=\frac{1}{\hbar^2}\frac{\partial^2 \epsilon_k}{\partial \vec{k}\partial \vec{k}}\frac{\partial}{\partial t}(\hbar \vec{k})
=\frac{1}{m^*_k}\frac{\partial}{\partial t}(\hbar \vec{k}) .
\eeq
The last equality is for the particular case of an isotropic band, with 
\beq
\frac{1}{m^*_k}\equiv\frac{1}{\hbar^2}\frac{\partial^2\epsilon_k}{\partial k^2} .
\eeq
According to semiclassical transport theory, in the presence of an external force $\vec{F}_{ext}^k$
\beq
\frac{\partial}{\partial t}(\hbar \vec{k}) = \vec{F}_{ext}^k .
\eeq
The total force exerted on a Bloch electron is
\beq
m_e\frac{d \vec{v}_k}{dt}\equiv\vec{F}_{tot}^k=\frac{m_e}{m^*_k} \vec{F}_{ext}^k =  \vec{F}_{ext}^k +
\vec{F}_{latt}^k
\eeq
with $m_e$ the bare electron mass, and $\vec{F}_{latt}^k$ the {\it force exerted by the lattice on the electron} of wavevector $k$, given by
\beq
\vec{F}_{latt}^k=(\frac{m_e}{m^*_k}-1) \vec{F}_{ext}^k
\eeq

Near the bottom of the band $m^*_k$ is positive and $\vec{F}_{latt}^k$ is small. Near the top of the band,
$m^*_k$ is negative and $\vec{F}_{latt}^k$ is larger than  $\vec{F}_{ext}^k$ and points in opposite direction, 
causing the electron near the top of the band to accelerate in direction $opposite$ to the external force.

When the lattice exerts a force on the electron, 
by Newton's third law the electron exerts a force on the lattice, or in other words transfers momentum
to the lattice. This indicates that the electrons that are most effective in transferring momentum 
from the electrons to the body are electrons near the top of the band. In other words, holes.
This will answer the  question of how the momentum of the supercurrent is transferred
to the body as a whole in a reversible way, without  energy dissipation. 

We consider conduction in crossed electric and magnetic fields as in the standard
Hall effect measurement.  It is easy to see that the total force exerted by the lattice on the carriers 
in direction perpendicular to the current flow is zero
 for a band close to empty with $R_H<0$ and is not zero for 
 a band close to full and $R_H>0$. The total force exerted by both the
 lattice and the external fields on the current carrying carriers in direction 
 perpendicular to their motion has to be zero, hence from  Eq. (21)
 \beq
\sum_{occ}\vec{F}_{tot}^k=\sum_{occ} \frac{m_e}{m^*_k} \vec{F}_{ext} ^k=0
\eeq
where the sum is over occupied $k$ states. 
 For the case $R_H<0$ and the band  close to empty we can assume that the effective mass
 is independent of $k$, $m^*_k=m^*$. From Eq. (23)
 \beq
 \sum_{occ} \frac{m_e}{m^*_k} \vec{F}_{ext}^k=
 \frac{m_e}{m^*} \sum_{occ} \vec{F}_{ext}^k=0
 \eeq
 therefore 
 \beq
  \sum_{occ} \vec{F}_{ext}^k=0
  \eeq
  and Eqs. (22), (24) and (25) imply
  \beq
 \sum_{occ}\vec{F}_{latt}^k=0
 \eeq
 so that the total force exerted by the lattice on the carriers in direction perpendicular to the 
 current flow is zero, and so is the total force exerted by the carriers on the lattice.
 
 Instead, for a band that is close to full and $R_H>0$, we cannot assume that 
 $m^*_k$ is independent of $k$ for the occupied states, instead we assume
 $m^*_k=-m^*$ for the empty states. Eq. (23) then implies       
 \beq
\sum_{occ}\vec{F}_{tot}^k=-\sum_{unocc}\frac{m_e}{m^*_k} \vec{F}_{ext} ^k= - \frac{m_e}{m^*}
\sum_{unocc} \vec{F}_{ext} ^k=0
\eeq
and from Eqs. (21)  and (27)
 \beqn
  & &     \sum_{occ}\vec{F}_{latt}^k=-\sum_{occ}\vec{F}_{ext}^k
  \\ \nonumber
  &=&
  -\sum_{all}\vec{F}_{ext}^k+\sum_{unocc}\vec{F}_{ext}^k  \\ \nonumber
  &=&  -\sum_{all}\vec{F}_{ext}^k=
  -2N e \vec{E} \neq 0  
  \eeqn
  where $N$ is the number of $k-$points in the Brillouin zone. To obtain Eq. (28) we used that
  the external force perpendicular to the current flow is
  \beq
  \vec{F}_{ext}^k=e\vec{E}+\frac{e}{c}\vec{v}_k\times\vec{B}
  \eeq
  with $\vec{E}, \vec{B}$ the electric and magnetic fields.

  Eq. (28) shows that when $R_H>0$ the lattice exerts a force on the conducting carriers that is
  perpendicular to the current flow.
  Conversely, the conducting carriers exert a force on the lattice or, in other words,
  transfer momentum to the lattice in direction perpendicular to the current flow. 
  In contrast, if the carriers are electrons with $R_H<0$ there is no net force exerted by electrons on the lattice nor by 
  the lattice on electrons 
  in direction perpendicular to the current flow, hence no momentum transfer from the carriers to the lattice.
 This is, in essence, why hole carriers are indispensable for superconductivity  \cite{momentum}.
 
   For the case of interest here, the superconductor-normal transition in the presence of a magnetic
  field, the electric field $\vec{E}$ above is the Faraday electric field parallel to the 
  normal-superconductor phase boundary that gets generated when the phase boundary
  moves due to change in magnetic flux. The flow of hole carriers tranferring momentum from
  electrons to the body is in the direction of motion of the phase boundary, perpendicular to
  the phase boundary.
 We have discussed elsewhere in quantitative detail how the momentum balance takes place in a cylindrical geometry 
 resulting in transfer of angular momentum between the current carriers and the body as a whole \cite{momentum}.
 The physics is the same in a linear geometry for a superconducting wire carrying a supercurrent
 entering the normal state. Here the momentum transferred is linear momentum instead of angular momentum. 
 We explain it qualitatively in Fig. 4.

       \begin{figure} [htb]
 \resizebox{8.5cm}{!}{\includegraphics[width=6cm]{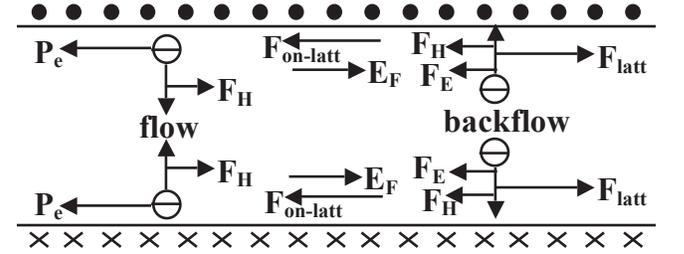}}
 \caption {Explanation of what happens to the mechanical momentum of the supercurrent
 when a superconducting wire carrying a supercurrent goes normal. Current flows to the right, electrons
 move to the left  carrying momentum $P_e$. Magnetic field points out of the paper at the top of the wire
 and into the paper at the bottom. As the wire  becomes normal magnetic field lines move towards the center
 of the wire together with the normal-superconductor phase boundary, and a Faraday electric field $E_F$ pointing to the right is generated. Electrons carrying the 
 supercurrent  flow inward
 towards the center of the wire (flow) and are stopped by the Lorentz force $F_H$, while other
 electrons further in (not shown) are accelerated by the Faraday field $E_F$. 
 At the same time, normal electrons right outside the normal-superconductor phase boundary
 flow outward (backflow). Equivalently, the backflow is carried by holes moving inward.
 The backflow electrons experience electric and magnetic Lorentz forces
 pointing to the left, balanced by the force $F_{latt}$ exerted on them by the lattice.
 At the same time, these electrons exert an equal and opposite force $F_{on-latt}$ on the lattice
 that transfers the momentum of the supercurrent to the body as a whole.
 }   
 \label{figure1}
 \end{figure} 
 
 The reverse process to the supercurrent stopping when a superconductor in a magnetic field goes normal
 is the Meissner effect, the process where a magnetic field is expelled from the interior of a normal metal
 when it becomes superconducting and a supercurrent develops. Here, one has to explain 
 how a supercurrent starts flowing in direction opposite to the direction dictated by the 
 Faraday electric field that develops as the magnetic field is expelled, as well as how
 momentum is conserved. The physics driving the effect is illustrated in Video 1 (Fig. 5).
 
        \begin{figure} [htb]
 \resizebox{8.5cm}{!}{\href{http://dx.doi.org/10.1117/12.2269644.1}{\includegraphics[width=6cm]{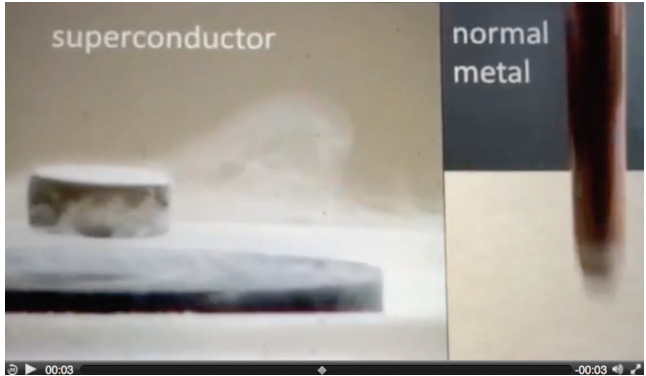}}}
 \caption { Video 1: illustration of the physics driving the Meissner effect within the theory discussed in
 this paper \cite{dyn2}. 
 In the left panel,  a magnet on top of a normal metal is lifted as the metal is cooled and becomes
 superconducting (Meissner effect). In the right panel, a magnet is lifted by the motion of a normal metal tube
 placed around it, due to the motion of the electrons in the metal tube and Faraday's law. We propose that
 the physics driving the lifting of the magnet in the left panel is the same as in the right panel: motion
 of electric charge and Faraday's law. The theory predicts that when a normal metal becomes
 superconducting, electrons are expelled from its interior to the surface, carrying
 magnetic field lines with it. {\href{http://dx.doi.org/10.1117/12.2269644.1}{http://dx.doi.org/10.1117/12.2269644.1}} } 
 \label{figure1}
 \end{figure}

 \section{hole superconductivity in materials}
 We have proposed a variety of models to describe superconductivity originating in pairing of hole carriers, initially motivated by the
 physics of high $T_c$ cuprates. 
We found that they describe in a  natural way several salient properties of cuprate superconductors
\cite{holebcs,marsiglio,holesc}, 
 in particular their:
  
  (i) Dome-like $T_c$ versus hole concentration dependence
  
  (ii) Positive pressure dependence of $T_c$
  
  (iii) Crossover between strong and weak coupling regimes as the hole concentration increases
  
  (iv) Crossover from incoherent to coherent behavior both as the hole concentration increases and as superconductivity
  sets in
  
  (v) Tunneling asymmetry, with larger current for negatively biased sample
  
  (vi) Apparent violation of conductivity sum rule, and transfer of optical spectral weight from high frequencies
  to low frequencies as superconductivity sets in.
  
  In addition, we have argued \cite{matmech} that these models lead to hole pairing and superconductivity in  the following classes \cite{classes}
   of superconducting materials:
  
  (1) Hole-doped cuprates  \cite{holebcs,marsiglio}
  
  (2) Electron-doped cuprates \cite{holesedoped}
  
  (3) Magnesium diboride   \cite{mgb2}
  
  (4) Transition metal series alloys \cite{tm}
  
  (5) Iron pnictides \cite{pnic}
  
  (6) Iron selenides  \cite{matmech}
  
  (7) Doped semiconductors  \cite{matmech,dopedsc}
  
  (8) Elements under high pressure \cite{hamlin,cava}
  
  (9) Sulphur hydride \cite{h2s}
  
  (10) A-15 materials \cite{a15,all}
  
  (11) All other superconductors \cite{all}
  
        \begin{figure} [htb]
 \resizebox{8.5cm}{!}{\includegraphics[width=6cm]{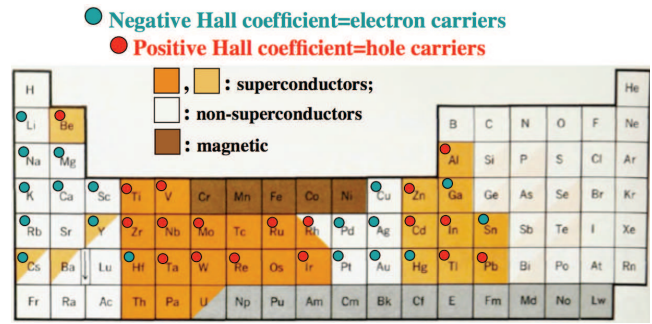}}
 \caption { Periodic table, showing the preponderance of superconductors among positive Hall
 coefficient elements and non-superconductors among negative Hall coefficient elements}
 \label{figure1}
 \end{figure} 

   For the simplest materials, the elements, there is an obvious preponderance of positive Hall coefficient 
  for superconducting elements and negative Hall coefficients for nonsuperconducting 
  elements \cite{h6,correlations}, as shown in Fig. 6.

 \section{discussion}

 . The evidence from materials that hole carriers are necessary for superconductivity
continues to accumulate. Of course sometimes it is the case that in a multiband situation electron
carriers exist and dominate the transport, in which case it may not be obvious that hole carriers
also exist and are responsible for superconductivity. For example, for a long time it was not 
known  that in electron-doped cuprates \cite{edoped} there are hole carriers and they dominate the transport
in the regime where they become superconducting \cite{greene}.
We predicted it in 1989 \cite{stanford}.
Similarly, we predict that in other materials where it is not obvious that hole carriers exist,
for example  the
very low carrier density n-doped semiconductor  $SrTiO_3$ \cite{sr1,srti}, hole
carriers will   eventually be found.

The Meissner effect is a universal property of superconductors. 
The reversible mechanism by which momentum is exchanged between the supercurrent and the body
as a whole in the Meissner effect and its reverse (superconductor-to-normal transition in a magnetic field) is undoubtedly universal for all superconductors.
We have shown that only hole carriers can do this in a reversible fashion.
Therefore, all superconductors are hole superconductors.
For any superconductor that appears to not have hole carriers, we predict that hole carriers will eventually be found. The search for new
 superconducting materials with high critical temperatures should focus on hole conductors
 and in particular materials where holes conduct through closely spaced negatively 
 charged anions \cite{matmech}.

\end{document}